\documentclass[twoside]{article}
\usepackage{fleqn,espcrc2}
\usepackage{epsfig,amsmath}

\begin{document}
\pagestyle{plain}
\title{Renormalization of the antisymmetric tensor field propagator and dynamical
generation of the $1^{+-}$ mesons in Resonance Chiral Theory\thanks{%
Presented by J.\ T. This work was supported in part by the Center for
Particle Physics (project no.\ LC 527), GACR (grant no.\ 202/07/P249) and by
the EU Contract No.\ MRTN-CT-2006-035482, \lq\lq\textsc{Flavia\textit{net}}%
''. }
\begin{picture}(0,0)(0,0)
\put(-38,65){\rm\small PSI-PR-08-14}
\put(-38,55){\rm\small PUPT-2283}
\end{picture}
}
\author{K.~Kampf\address{Paul Scherrer Institut, Ch-5232 Villigen PSI, Switzerland}%
\address{IPNP, MFF, Charles University, V~Hole\v{s}ovi\v{c}k\'{a}ch 2,
CZ-180 00 Prague 8, Czech Republic}, J.~Novotn\'y$^{\mathrm{b}}$,
J.
Trnka$^{\mathrm{b}}$%
\address{Department of Physics, Princeton University,
        08540 Princeton, New Jersey, USA},}

\begin{abstract}
We discuss the renormalization of the $1^{--}$ vector meson
propagator within Resonance chiral theory at one loop. Using
the particular form of the interaction Lagrangian we show that
additional poles of the renormalized propagator corresponding to
$1^{+-}$ degrees of freedom can be generated. We give a concrete
example of such an effect.
\end{abstract}

\maketitle

\section{Introduction}

Non-perturbative investigation of the QCD dynamics in the low energy region
by means of the effective Lagrangian approach has made considerable
progress recently. In the very low energy region ($E\ll \Lambda _H\sim 1\,GeV$%
)
the chiral perturbation theory ($\chi $PT)
\cite{Weinberg,Gasser1,Gasser2} based on the spontaneously broken
chiral symmetry $SU(3)_L\times SU(3)_R$ grew into a very
successful model-independent tool for description of the Green
functions (GF) of quark currents and related low-energy
phenomenology. $\chi $PT is organized as a rigorously defined
simultaneous perturbative expansion in small momenta and light
quark masses. Recent calculations are
performed at the next-to-next-to-leading order $O(p^6)$ \cite{Bijnens:2006zp}%
.

In the intermediate energy region ($\Lambda _H\leq E<2\,GeV$),
however, the situation is less satisfactory. The set of relevant
degrees of freedom includes now the low lying resonances and
because there is no mass gap existing in the spectrum, the
effective theory in this region cannot be constructed as a
straightforward extension of the $\chi $PT low energy expansion.
On the other hand, the considerations based on the large $N_C$
expansions together with high-energy constraints derived from
perturbative QCD and operator product expansion (OPE) allow to introduce another type of
effective Lagrangian description, corresponding to the leading
order in $1/N_C$ and reflecting the basic features of QCD in the
$N_C\rightarrow \infty $ limit. Namely, the spectrum
consisting of an infinite tower of free stable mesonic resonances
exchanged in each channel requires infinite number of resonance
fields in the $U(3)_L\times U(3)_R$ symmetric Lagrangian with
interaction
vertices suppressed by an appropriate power of $\,N_C^{-1/2}$and (since the $%
1/N_C$ expansion is correlated with semiclassical expansion) only tree
graphs have to be taken into account in the leading order. An approximation
to this general picture consisting in limiting the number of resonance field
to one in each channel and matching the resulting theory in the high energy
region with OPE is known as Resonance Chiral Theory (R$\chi $T) \cite
{Ecker1,Ecker2}. Integrating out the resonance fields from the Lagrangian of
R$\chi $T in the low energy region and subsequent matching with $\chi $PT
has become a very successful tool for estimates of the resonance contribution
to the values of the $O(p^4)$ \cite{Ecker1} and $O(p^6)$ \cite
{Cirigliano:2006hb,Kampf:2006bn} low energy constants (LEC) entering the $%
\chi $PT Lagrangian.

Though the usual chiral power counting fails within R$\chi $T
due to the presence of an additional heavy scale (the mass of the
resonances) and the usual Weinberg formula \cite{Weinberg} cannot
be generalized here (because of the lack of
a scale playing the role analogous to
$\Lambda _H$), it seems to be fully legitimate to go beyond the
tree level R$\chi $T and calculate the loops \cite{vecform}.
Being suppressed by one power of $1/N_C$, the loops allow to
encompass such NLO effects in the $1/N_C$ expansion as resonance
widths
and final state interaction and
to determine the NLO resonance contribution to LEC (and their
running with the renormalization scale).

However, we have to be ready for both technical and conceptual complications
connected with renormalization of the effective theory for which no natural
organization of the expansion (other than the $1/N_C$ counting) exists.
Especially, because there is no natural analog of the Weinberg power
counting in R$\chi $T, we can expect mixing of the naive chiral orders in
the process of the renormalization (\emph{e.g} the loops renormalize the $%
O(p^2)$ LEC and also counterterms of unusually high chiral orders are
needed). Also, lack of appropriate protective symmetry can bring about
appearance of new poles in the GF corresponding to new degrees of freedom%
\footnote{%
Such a phenomenon within effective theory has been observed in
another context \emph{e.g.} in \cite{Wu:2008rr}} which are frozen
at the tree level. The latter might be felt as a pathological
artefact of the not carefully enough formulated theory,
particularly because this extra poles might be negative norm ghost
or tachyons \cite{Slovak}. On the other hand, however, we could
also try to take advantage of this feature and adjust the poles in
such a way that they correspond to the well established resonance
states.

In the following we will illustrate these problems in more detail.
As an explicit example we use the one-loop renormalization of the
propagator
corresponding to the antisymmetric tensor field which originally describes
the $1^{--}$ vector resonance ($\rho $ meson) at the tree level. We
will show that the loop corrections to the propagator could lead
to the dynamical generation of various types of $1^{--}$ and
$1^{+-\text{ }}$ states and that the appropriate adjustment of
coupling constants allows us to generate in this way the one which
could be identified with the $b_1(1235)$ meson. The details of the
calculations and further discussion will be provided in
\cite{clanek}.

\section{Antisymmetric tensor field propagator}

The $1^{--}$ resonance part of the R$\chi $T Lagrangian within the
antisymmetric tensor formalism reads \cite{Ecker1,Ecker2}
\begin{equation}
\mathcal{L}=-\frac 14\langle D^\mu R_{\mu \nu }D_\alpha R^{\alpha \nu
}\rangle +\frac 12M^2\langle R^{\mu \nu }R_{\mu \nu }\rangle +\mathcal{L}%
_{int},  \label{lagrangian}
\end{equation}
where $R_{\mu \nu }=R_{\mu \nu }^aT^a$ with normalized $U(3)$ generators $%
T^a $ , $R_{\mu \nu }^a$ are antisymmetric tensor fields with appropriate
quantum numbers and $D_\alpha
$ is the
usual chiral covariant derivative. $\mathcal{L}_{int}$ is the interaction
Lagrangian
which will be specified later. The full antisymmetric tensor field
propagator $\delta ^{ab}\Delta _{\mu \nu \alpha \beta
}(p)=-\mathrm{i}F.T.\langle 0|TR_{\mu \nu }^a(x)R_{\alpha \beta
}^b(0)|0\rangle$ has  in  general  the following  tensor structure
\[
\Delta _{\mu \nu \alpha \beta }(p)=-2\Pi _{\mu \nu \alpha \beta }^L\Delta
^L(p^2)+2\Pi _{\mu \nu \alpha \beta }^T\Delta ^T(p^2),
\]
where $\Pi _{\mu \nu \alpha \beta }^{L,T}$ are longitudinal and transverse
projectors ($P_{\mu \nu }^T=g_{\mu \nu }-p_\mu p_\nu/p^2$)
\begin{eqnarray*}
\Pi _{\mu \nu \alpha \beta }^T &=&\frac 12\left( P_{\mu \alpha }^TP_{\nu
\beta }^T-P_{\nu \alpha }^TP_{\mu \beta }^T\right), \\
\Pi _{\mu \nu \alpha \beta }^L &=&\frac 12\left( g_{\mu \alpha }g_{\nu \beta
}-g_{\nu \alpha }g_{\mu \beta }\right) -\Pi _{\mu \nu \alpha \beta }^T.
\end{eqnarray*}
Note that, for $p^2=m^2>0$ we can express $\Pi _{\mu \nu \alpha \beta
}^{L,T} $ as the polarization sums
\begin{eqnarray*}
-2\Pi _{\mu \nu \alpha \beta }^L &=&\sum_\lambda u_{\mu \nu }^{(\lambda
)}(p)u_{\alpha \beta }^{(\lambda )}(p)^{*}, \\
2\Pi _{\mu \nu \alpha \beta }^T &=&\sum_\lambda w_{\mu \nu }^{(\lambda
)}(p)w_{\alpha \beta }^{(\lambda )}(p)^{*},
\end{eqnarray*}
where
\begin{eqnarray}
u_{\mu \nu }^{(\lambda )}(p) &=&\frac{\mathrm{i}}m\left( p_\mu \varepsilon
_\nu ^{(\lambda )}(p)-p_\nu \varepsilon _\mu ^{(\lambda )}(p)\right)
\label{wave_function}, \\
w_{\mu \nu }^{(\lambda )}(p) &=&\widetilde{u}_{\mu \nu }^{(\lambda
)}(p)=\frac 12\varepsilon _{\mu \nu \alpha \beta }u^{(\lambda )\alpha \beta
}(p)
\end{eqnarray}
and $\varepsilon _\nu ^{(\lambda )}(p)$ are the usual spin-one polarization
vectors with mass $m$. The possible poles $p^2=m_{L,T}^2>0$ of $\Delta
^L(p^2)$ and $\Delta ^T(p^2)$ correspond therefore both to the spin-one
states which couple to the fields $\partial ^\alpha R_{\alpha \mu }^a$ and $%
\varepsilon ^{\mu \nu \alpha \beta }\partial_\nu R^{a}_{\alpha \beta }$
respectively and have therefore the same quantum numbers up to the parity.

At LO in the $1/N_C$ expansion the Lagrangian (\ref{lagrangian}) gives $%
\Delta _{LO}^L(p^2)=1/(p^2-M^2)$ and $\Delta _{LO}^T(p^2)=M^{-2}$ so that
the $1^{--}$ resonance multiplet appears as a pole in $\Delta ^L(p^2)$.
Because there is no pole in $\Delta ^T(p^2)$ at this order, no additional $%
1^{+-}$ state is propagated. Beyond LO we get generally
\begin{eqnarray*}
\Delta ^L(p^2)^{-1} &=&p^2-M^2-\Sigma _L(p^2), \\
\Delta ^T(p^2)^{-1} &=&M^2+\Sigma _T(p^2),
\end{eqnarray*}
where the self-energies $\Sigma_{L,T}(p^2)$ are of the order $1/N_C$ at
least. In the next section we present the results of the calculation of the
renormalized self-energies $\Sigma ^{L,T}(p^2)$ in the chiral limit at NLO
for a concrete form of the interaction Lagrangian $\mathcal{L}_{int}$. A more
systematic treatment will be given in \cite{clanek}.

\section{The one-loop self-energies within R$\chi $T}

In what follows we limit ourselves to the interaction Lagrangian $\mathcal{L}%
_{int}$ with at most two derivatives and up to two resonance fields. Writing
explicitly only those terms that contribute to the one-loop self-energies we
have \cite{Ecker1}, \cite{VVP}
\[
\mathcal{L}_{int}=\frac{\mathrm{i}G_V}{2\sqrt{2}}\langle R^{\mu \nu }[u_\mu
,u_\nu ]\rangle +2d_1\langle D_\beta u^\sigma \{\widetilde{R}_{\alpha \sigma
},R^{\alpha \beta }\}\rangle
\]
\[
+2d_3\langle u^\lambda \{D_\nu R^{\mu \nu },\widetilde{R}_{\mu \lambda
}\}\rangle +2d_4\langle u_\nu \{D^\alpha R^{\mu \nu },\widetilde{R}_{\mu
\alpha }\}\rangle,
\]
where $\widetilde{R}_{\alpha \beta }=\varepsilon _{\alpha \beta \mu \nu
}R^{\mu \nu }/2$. In the large $N_C$ limit the couplings are $%
G_V=O(N_C^{1/2})$ and $d_i=O(1)$ and apparently the intrinsic
parity odd part is of higher order. However, the trilinear
vertices contributing to the one-loop self-energies are
$O(N_C^{-1/2})$ in both cases due to the appropriate power of
$1/F=O(N_C^{-1/2})$ accompanying $u_\alpha $. Therefore the
operators with two resonance fields cannot be eliminated using the
large $N_C$ arguments. Also nonzero $d_i$ are required in order to
satisfy the OPE constraints for VVP GF at the LO \cite{VVP}.

In order to cancel the infinite part of the one-loop self-energies we have
to introduce a set of counterterms. Because the interaction terms are $O(p^2)$
we would expect (by the analogy with $\chi $PT power counting) these
counterterms to have four derivatives at most. However, the nontrivial
structure of the free resonance propagator (namely the presence of the $%
\Delta _{LO}^T$ $\Pi _{\mu \nu \alpha \beta }^T$ part) results in
the failure of this naive expectation. In fact we need
counterterms with up to six derivatives, namely
\begin{equation}
\mathcal{L}_{ct}=\mathcal{L}_{ct}^{(0)}+\mathcal{L}_{ct}^{(2)}+\mathcal{L}%
_{ct}^{(4)}+\mathcal{L}_{ct}^{(6)}\,.  \label{ct}
\end{equation}
The complete list  of the counterterms and their infinite parts is
postponed to \cite{clanek}.
Let us only note that $\mathcal{L}_{ct}^{(2)}$ contains a new type
of kinetic term $\frac Y4\langle D_\alpha R^{{\mu }{\nu }}D^\alpha
R_{{\mu }{\nu }}\rangle$. Provided such a term was included in the
LO Lagrangian from the
very beginning, the propagator would have an additional pole in $\Delta _{LO}^T$%
. However, interpretation of such a pole as a $1^{+-}$ state would
be problematic. According to the sign of $Y$ this state would be
either a tachyon or a negative norm ghost \cite{Slovak}.

Evaluating the one-loop Feynman graphs
and adding the polynomial contributions from the counterterms
(\ref {ct}) we get the $\chi $PT minimally subtracted
self-energies $\Sigma _{L,T}^r(p^2)$. The equation for the poles
of $\Delta ^L(p^2)$ has then an approximative perturbative solution
$p^2=M^2+\Sigma _L^r(M^2)$ corresponding to the original $1^{--}$
vector resonance with LO mass $M^2$, which develops a mass
correction and a finite width of the order $O(1/N_C)$ due to the
loops and which we identify with the $\rho$ meson. This allows to
re-parameterize perturbatively $\Sigma _{L,T}(s)$ in terms of $M_\rho $ and $%
\Gamma _\rho $ and requiring further $M^2=M_\rho ^2$ we get for
$\sigma _{L,T}^r(x)=$ $\Sigma _{L,T}^r(xM_\rho ^2)/M_\rho ^2$
\cite{clanek}
\begin{eqnarray*}
{\sigma }_L^r(x) &=&\frac 1\pi \frac{\Gamma _\rho }{M_\rho }\left[ 1-x^2%
\widehat{B}(x)+\sum_{i=1}^3a_i(x^i-1)\right] \\
&&-\frac{40}9\left( \frac{M_\rho }{4\pi F_\pi }\right) ^2d_3^2(x^2-1)^2%
\widehat{J}(x) \\
{\sigma }_T^r(x) &=&\frac 1\pi \frac{\Gamma _\rho }{M_\rho }%
\sum_{i=0}^3b_ix^i+\frac{20}9\left( \frac{M_\rho }{4\pi F_\pi }\right)
^2d_3^2
\end{eqnarray*}
\[
\times \left( 2+(1+6\gamma +\gamma ^2)x+2\gamma ^2x^2\right) (x-1)^2\widehat{%
J}(x).
\]
Here we put for further numerical estimates $F=F_\pi $, $\gamma
=d_4/d_3\sim O(1)$ (for $d_3$ we take the value from \cite{VVP}) and we
have introduced the re-scaled free parameters $a_i$ and $b_i$ with
natural size $O(1)$ in the large $N_C$ expansion. These can be
expressed in terms of renormalization scale independent
combinations of the renormalized counterterms couplings and $\chi
$logs. The loop functions $\widehat{B}(x)$ and $\widehat{J}(x)$
are given on the first sheet as
\begin{eqnarray}
\widehat{B}(x) &=&1-\ln (-x) ,  \nonumber \\
\widehat{J}(x) &=&x^{-1}\left[ 1-\left( 1-x^{-1}\right) \ln (1-x)\right] ,
\label{loop_functions}
\end{eqnarray}
where we take the principal branch of the logarithm ($-\pi <\mathrm{{Im}\ln
x\leq \pi }$) with a cut for $x<0$. On the second sheet we have then $\widehat{%
B}^{II}(x-\mathrm{i}0)=\widehat{B}^I(x+\mathrm{i}0)
=\widehat{B}^I(x-\mathrm{i}0)+2\pi \mathrm{i} $ and similarly for
$\widehat{J}^{II}(x)
$. 

\begin{figure}[tb]
\epsfxsize=6.5cm\epsfbox{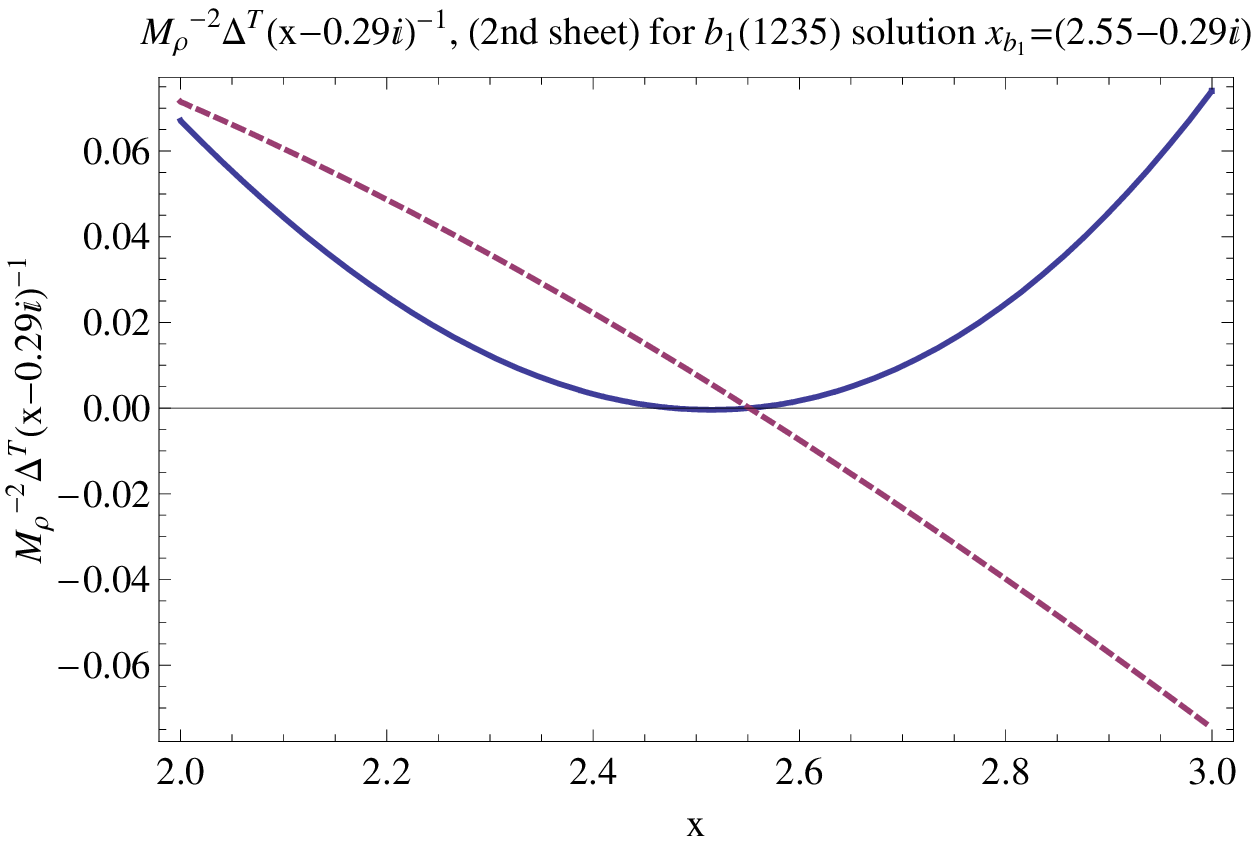}
\put(-158,18){\epsfig{width=2.48cm,figure=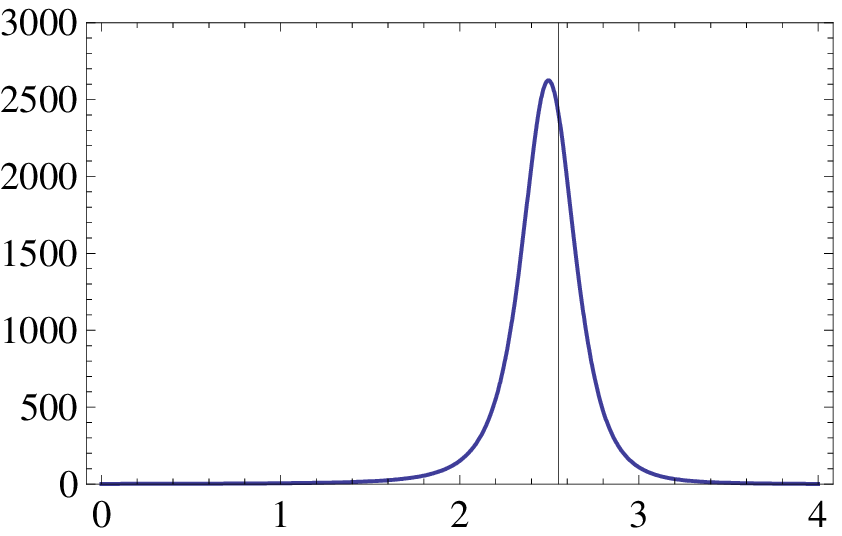}}\vspace{-1cm}
\caption{The real (full line)and imaginary part (dots) of the
function $1/(M_\rho^2\Delta^T(p^2))$ for the section $p^2/M_\rho
^2=x-iM_{b_1}\Gamma _{b_1}/M_\rho ^2$ on the second sheet. The
$b_1(1235)$ pole corresponds to zero at
$x_{b_1}=2.55-{\rm{i}}0.29$. The shape of $|\Delta _L(p^2)|^2$ on
the first sheet for $p^2$ real is in the small frame.
\vspace{-0.8cm}} \label{figo}
\end{figure}

The equation for the poles of $\Delta ^T(p^2)$ has only
non-perturbative solutions of the order $O(N_C)$. The $1^{+-}$
states corresponding to them therefore   decouple in the
$N_C\rightarrow \infty $ limit, however, for physical values of
$M_\rho $, $\Gamma _\rho $ and $F_\pi $ (and for reasonable $O(1)$
values of the parameters $b_i$ and $\gamma $), the position of
poles can lie well within the intermediate energy region we are
interested in. The nature of the corresponding states, which is
also controlled by the free parameters $b_i$ and $\gamma $,  is
rich and covers bound states $0<p^2<M_\rho ^2$, (which also might
be negative norm ghosts), tachyonic poles $p^2<0$, resonance poles
in the lower complex half-plane on the second sheet or even
complex conjugated pair of Lee-Wick poles on the
first sheet.  It is not straightforward to formulate general conditions for $%
b_i$ and $\gamma $ under which there is \emph{no} pole in $\Delta
^T(p^2)$ at all, on the other hand we can rather easily  arrange
them to obtain  the pole corresponding \emph{e.g.} to the
$b_1(1235)$ meson. This can be
achieved in many ways \emph{e.g. }for the choice $b_0=-b_1=-3.52$, $b_2=-3.07$%
, $b_3=1$ and $\gamma =-0.52$. The plot of the denominator of
$\Delta^T(p^2)$ for this particular choice for the section
$p^2/M_\rho ^2=x-iM_{b_1}\Gamma _{b_1}/M_\rho ^2$ on the second
sheet and  the shape of $|\Delta _L(p^2)|^2$ on the first sheet
for $p^2/M_\rho ^2$ real are depicted  in Fig~\ref{figo}.

\section{Conclusion}
We have illustrated the problems connected with loop calculations
within R$\chi$T using the one loop renormalization of the
propagator of antisymmetric tensor field which describes $1^{--}$
resonance multiplet at the leading order of the large $N_C$
expansion as a concrete example. We have found that new $1^{+-}$
states of various nature (including pathological ones like
negative norm ghosts and tachyons) can be dynamically generated.
For example, for a wide range of parameters $b_i$, one such a state
can be identified with $b_1(1235)$ meson. Further discussion of
possible applications of this feature is postponed to
\cite{clanek}.

\end{document}